# WKB approach for structured waveguides

*M.I. Ayzatsky*
*National Science Center Kharkov Institute of Physics and Technology (NSC KIPT), 61108, Kharkov, Ukraine*
*E-mail: mykola.aizatsky@gmail.com*

The results of the development of an approximate approach, which can be considered as an analogue of the WKB method, are presented. This approach gives possibility to divide the electromagnetic field in structured waveguides into forward and backward components and simplify the analysis of the field characteristics, especially the phase distribution. The accuracy of this method was estimated by comparing the solution of the approximate system of equations with the solution of the general system of equations. For this, a special code was written that combines the proposed approach with the more accurate one developed earlier. The results of this comparison showed that the accuracy of the proposed approximate approach for sections with "group velocity" gradients, similar to the SLAC section, is about 1%.

## 1. INTRODUCTION

Structured waveguides[1] based on coupled resonators play important role in many applications (accelerators, RF devices). There are powerful codes that can be used to calculation RF characteristic of structured waveguides, but it is difficult to use them for investigation properties and making preliminary design. For these purposes the fast and versatile models are needed.

Moreover, there is a question about the accuracy of approximate models which widely used for calculating the characteristics of structured waveguides with a sufficiently slow change in their dimensions. This is especially important when studying the energy spread and its minimum, which can be achieved using a low-level RF feedback system (see, for example, [1,2,3,4,5,6,7,8,9]). Indeed, the energy spread caused by transients arising when the beam turn on can be decreased by choosing the injection time, the profiles of RF amplitude and phase at the section RF input (input coupler), and, perhaps, the profile of beam current. There are many works with results of study these processes. All these results were obtained on the basis of some equations describing the excitation of a waveguide by an external current. There is full set of equations in the case of homogeneous periodic waveguides. In the case of inhomogeneous waveguides, the approximate models are used. Nobody knows the correctness of these models as they are based on the physical, not mathematical suggestions and manipulations. Thus, in addition to basic models, approximate approaches with a known accuracy are needed.

In this paper we present the results of developing on the base of the CASCIE[2] code [10] an approximate approach which can be considered as an analog of the WKB method

## 2. BASIC EQUATIONS

Let us consider a cylindrical waveguide with the z-axis coinciding with the symmetry axis of a cylinder. We will consider only axially symmetric fields with $E_z, E_r, H_\varphi$ components (TM fields). Time dependence is $\exp(-i\omega t)$. If we introduce arbitrary cross sections that are perpendicular to the z-axis and represent the tangential components of electric fields on these cross sections as the sum of basis functions

$$E_\tau^{(k)}(r) = \sum_s Q_s^{(k)} \varphi_s^{(k)}(r), \quad . \tag{1}$$

we can obtain such matrix equation

$$T^{(k)} Q^{(k)} = T^{+(k)} Q^{(k+1)} + T^{-(k)} Q^{(k-1)}, \tag{2}$$

where $T^{(k)}, T^{+(k)}, T^{-(k)}$ are the matrices and $Q^{(k)} = \left(Q_1^{(k)}, Q_2^{(k)}, Q_3^{(k)}, ....\right)^T$ are the vectors.

Indeed, we can represent the fields between two cross sections ($k^{th}$ and $(k+1)^{th}$) as an expansion with the short-circuit resonant cavity modes

$$\vec{E}^{(k)} = \sum_q e_q^{(k)} \vec{E}_q^{(k)}(\vec{r}), \tag{3}$$

$$\vec{H}^{(l)} = i \sum_q h_q^{(l)} \vec{H}_q^{(l)}(\vec{r}), \tag{4}$$

where $q = \{0, m, n\}$, $\vec{E}_q^{(l)}, \vec{H}_q^{(l)}$ are the solutions of homogenous Maxwell equations

---

[1] Waveguides that consist of similar (but not always identical) cells
[2] CASCIE - Code for Accelerating Structures - Coupled Integral Equations

$$rot\,\vec{E}_q^{(k)} = i\omega_q^{(k)}\mu_0\vec{H}_q^{(k)},$$
$$rot\,\vec{H}_q^{(k)} = -i\omega_q^{(k)}\varepsilon_0\vec{E}_q^{(k)} \quad (5)$$

with boundary condition $\vec{E}_\tau = 0$ on the metal surface

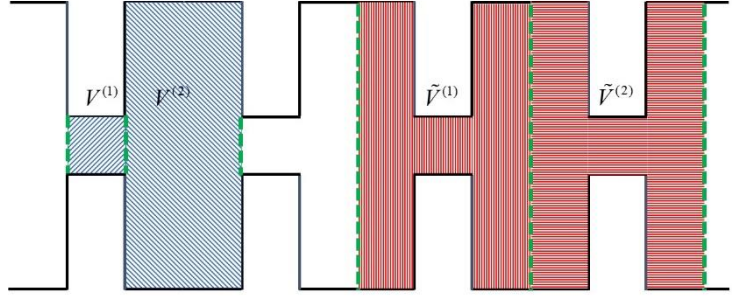

Figure 1

Under time dependence $\exp(-i\omega t)$ the equations for the amplitudes $e_{q,\omega}^{(k)}$ and $h_{q,\omega}^{(k)}$ can be written as

$$\left(\omega_q^{(k)2} - \omega^2\right)e_{q,\omega}^{(k)} = \left(\int_{S_k}[\vec{E}_c^{(k)}\vec{H}_q^{(k)*}]d\vec{S} + \int_{S_{k+1}}[\vec{E}_c^{(k+1)}\vec{H}_q^{(k)*}]d\vec{S}\right) = i\sum_s\left(-Q_s^{(k)}\Upsilon_{q,s}^{(k,k)} + Q_s^{(k+1)}\Upsilon_{q,s}^{(k+1,k)}\right),$$
$$h_{q,\omega}^{(k)} = -i\frac{\omega}{\omega_q^{(k)}}e_{q,\omega}^{(k)}. \quad (6)$$

where $\Upsilon_{q,s}^{(k,l)} = \frac{2\pi\omega_q^{(l)}}{N_q^{(l)}}\int_{S_k}\varphi_s^{(k)}(r)H_{q,\varphi}^{(l)*}(r)rdr$.

Equating the magnetic fields on the cross section between the $k^{th}$ and $(k-1)^{th}$ volumes

$$\sum_q\frac{\omega}{\omega_q^{(k)}\left(\omega_q^{(k)2} - \omega^2\right)}\sum_s\left(-Q_s^{(k)}\Upsilon_{q,s}^{(k,k)} + Q_s^{(k+1)}\Upsilon_{q,s}^{(k+1,k)}\right)H_{q,\varphi}^{(k)}(r) =$$
$$\sum_q\frac{\omega}{\omega_q^{(k-1)}\left(\omega_q^{(k-1)2} - \omega^2\right)}\sum_s\left(-Q_s^{(k-1)}\Upsilon_{q,s}^{(k-1,k-1)} + Q_s^{(k)}\Upsilon_{q,s}^{(k,k-1)}\right)H_{q,\varphi}^{(k-1)}(r) \quad (7)$$

and using the standard procedure of the Moment Method, we get the equation (2). The eigen vectors $\vec{E}_q^{(l)}$, $\vec{H}_q^{(l)}$ are known only for simplest resonators. This leads to the need to choose dividing cross sections separating significantly different, but geometrically simple volumes (see Figure 1, $V^{(1)}$, $V^{(2)}$). There the field singularities are often located and we need to use many basis functions in the expansion (1). With such choice the given above method is equivalent to the Coupled Integral Equations Method (CIEM) [11]. In this case the matrices $T^{+(k)}, T^{-(k)}$ can change significantly with changing the index $k$ even for the homogeneous waveguide (the equation (2) has a biperiodic symmetry).

Modified Coupled Integral Equations Method (MCIEM) [12] gives possibility to choose locations of dividing cross sections in places (see Figure 1, $\tilde{V}^{(1)}$, $\tilde{V}^{(2)}$), where the field has no singularities. The system of coupled integral equations in this approach can be formulated for longitudinal electrical fields in contrast to the standard approach where the transverse electrical fields are unknowns. The MCIEM (CIEM) can be effectively used to calculate characteristics of simple structured waveguides which have the analytic eigenwaves in the subdomains.

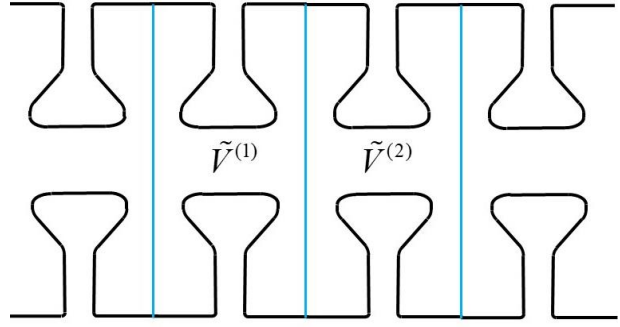

Figure 2

For more complicated waveguides (Figure 2) the given above procedure can be effectively realized numerically since many codes to calculate $\vec{E}_q^{(l)}$, $\vec{H}_q^{(l)}$ are developed now (SUPERFISH, HFSS, CST and others). In this case, it is possible to choose dividing cross sections in places where there are no field singularities or the field have simple structure (see Figure 1 and Figure 2, $\tilde{V}^{(1)}, \tilde{V}^{(2)}$)

We will be based on the MCIEM and use the possibilities of the CASCIE code [10] for computing $T^{(k)}, T^{+(k)}, T^{-(k)}$ matrices and to compare the approximate results with the results obtained from solving the equation (2).

Table 1 Determinant and eigen values of the matrix $\tilde{T}$

| Size of matrix $\tilde{T}$ | 2×2 | 3×3 | 4×4 | 5×5 | 6×6 |
|---|---|---|---|---|---|
| Determinant | -1.97E-05 | -4.41E-13 | -1.36E-23 | -5.97E-37 | -3.96E-53 |
| $\theta_1$ | -1.01E+00 | -1.00E+00 | -1.00E+00 | -1.00E+00 | -1.00E+00 |
| $\theta_2$ | 1.95E-05 | 3.98E-05 | 4.67E-05 | 4.78E-05 | 4.79E-05 |
| $\theta_3$ | | 1.11E-08 | 4.44E-08 | 7.59E-08 | 8.91E-08 |
| $\theta_4$ | | | 6.55E-12 | 4.22E-11 | 1.10E-10 |
| $\theta_5$ | | | | 3.89E-15 | 3.49E-14 |
| $\theta_6$ | | | | | 3.06E-18 |

Before proceeding with the transformation of the equation (2), we need to characterize the properties of matrices $T^{(k)}, T^{+(k)}, T^{-(k)}$. The matrices $T^{(k)}$ do not have any peculiarities (the determinants are not small and the condition numbers are not large) and we can rewrite the equation (2) as

$$Q^{(k)} = \tilde{T}^{+(k)} Q^{(k+1)} + \tilde{T}^{-(k)} Q^{(k-1)}, \quad (8)$$

where $\tilde{T}^{\pm(k)} = T^{(k)-1} T^{\pm(k)}$.

One of the important properties of the matrices $\tilde{T}^{\pm(k)}$ is its non-defectivity, except for a certain frequency range near the ends of the passbands.

The equation (8) describes the electric field at the given frequency $\omega$. It is known that for a homogeneous periodic waveguide ($\tilde{T}^{+(k)} = \tilde{T}^{-(k)} = \tilde{T}$) the electric field is the sum of an infinite number of eigen waves, which can differ greatly from each other. There are the propagation waves and the evanescent fields. Floquet multiplies $\lambda_s$ of the evanescent fields take both very small and very large values. If we designate the eigen values of the matrix $\tilde{T}$ as $\theta_s$, then it follows from (8) that the Floquet multiplies $\lambda_s$ of the electromagnetic field are determined by the characteristic equation

$$\lambda_s^2 - \theta_s^{-1} \lambda_s + 1 = 0, \quad (9)$$

from which we get

$$\lambda_s = 0.5 \theta_s^{-1} \mp \sqrt{0.25 \theta_s^{-2} - 1}, \quad (10)$$

To get a set of increasing and decreasing values of $\lambda_s$, the values of $\theta_s$ must decrease as the index $s$ increases. The same is true for inhomogeneous waveguides. Therefore, the determinants of matrices $\tilde{T}^{\pm(k)}$ will decrease if we increase their sizes and can become very small. It can lead to large errors when performing matrix operations. All of the above is confirmed by the results of calculations which is presented in Table 1. These data refer to a homogeneous disk-loaded waveguide with $a = 1.381$ cm, $b = 4.16186$ cm, $d = 2.9147$ cm, $t = 0.5842$ cm, $\varphi = 2\pi/3$, $f = 2.856$ GHz

In work [12], where the matrices have small dimensions (2×2), equation (2) was transformed by multiplying by the matrix $T^{-(k)-1}$. For the general case the matrices have larger dimensions (in our consideration 4×4 or more), their determinants become very small, and such transformation cannot be done. Therefore, it is necessary to change the procedure for obtaining the general WKB equations. This procedure should not contain the inverse matrices $\tilde{T}^{\pm(k)-1}$.

## 2. TRANSFORMATION OF THE BASIC EQUATIONS

As the equation (8) is of the second order, we can represent the solution of this matrix difference equation as the sum of two new vectors [13]

$$Q^{(k)} = Q^{(k,1)} + Q^{(k,2)}. \tag{11}$$

By introducing new unknowns $Q^{(k,i)}$ instead of the one $Q^{(k)}$, we can impose an additional condition. This condition we write in the form

$$Q^{(k+1)} = M^{r(k,1)}Q^{(k,1)} + M^{r(k,2)}Q^{(k,2)}, \tag{12}$$

where $M^{(k,i)}$ are arbitrary nondefecting matrices.

Using (11) and (12) we can rewrite (8) as

$$\tilde{T}^{+(k+1)}\left(M^{r(k+1,1)} - M^{r(k+1,2)}\right)Q^{(k+1,1)} = -\left(\tilde{T}^{-(k+1)} + \left(\tilde{T}^{+(k+1)}M^{r(k+1,2)} - I\right)M^{r(k,1)}\right)Q^{(k,1)} -$$
$$\left(\tilde{T}^{-(k+1)} + \left(\tilde{T}^{+(k+1)}M^{r(k+1,2)} - I\right)M^{r(k+1,2)} + \left(\tilde{T}^{+(k+1)}M^{r(k+1,2)} - I\right)\left(M^{r(k,2)} - M^{r(k+1,2)}\right)\right)Q^{(k,2)},$$
$$\tilde{T}^{+(k+1)}\left(M^{r(k+1,1)} - M^{r(k+1,2)}\right)Q^{(k+1,2)} = \left(\tilde{T}^{-(k+1)} + \left(\tilde{T}^{+(k+1)}M^{r(k+1,1)} - I\right)M^{r(k,2)}\right)Q^{(k,2)} +$$
$$\left(\tilde{T}^{-(k+1)} + \left(\tilde{T}^{+(k+1)}M^{r(k+1,1)} - I\right)M^{r(k+1,1)} + \left(\tilde{T}^{+(k+1)}M^{r(k+1,1)} - I\right)\left(M^{r(k,1)} - M^{r(k+1,1)}\right)\right)Q^{(k,1)}. \tag{13}$$

We choose the matrices $M^{(k,i)}$ ($i = 1, 2$) so that they satisfy quadratic matrix equations

$$\tilde{T}^{-(k+1)} + \left(\tilde{T}^{+(k+1)}M^{r(k+1,1)} - I\right)M^{r(k+1,1)} = 0,$$
$$\tilde{T}^{-(k+1)} + \left(\tilde{T}^{+(k+1)}M^{r(k+1,2)} - I\right)M^{r(k+1,2)} = 0. \tag{14}$$

Then the system (13) is transformed into

$$\tilde{T}^{+(k+1)}\left(M^{r(k+1,1)} - M^{r(k+1,2)}\right)Q^{(k+1,1)} =$$
$$\left\{\tilde{T}^{-(k+1)}M^{r(k+1,2)-1}\left(M^{r(k+1,1)} - M^{r(k+1,2)}\right) + \tilde{T}^{-(k+1)}M^{r(k+1,2)-1}\left(M^{r(k,1)} - M^{r(k+1,1)}\right)\right\}Q^{(k,1)} +$$
$$\tilde{T}^{-(k+1)}M^{r(k+1,2)-1}\left(M^{r(k,2)} - M^{r(k+1,2)}\right)Q^{(k,2)},$$
$$\tilde{T}^{+(k+1)}\left(M^{r(k+1,1)} - M^{r(k+1,2)}\right)Q^{(k+1,2)} = \tag{15}$$
$$-\left\{\tilde{T}^{-(k+1)}M^{r(k+1,1)-1}\left(M^{r(k+1,2)} - M^{r(k+1,1)}\right) + \tilde{T}^{-(k+1)}M^{r(k+1,1)-1}\left(M^{r(k,2)} - M^{r(k+1,2)}\right)\right\}Q^{(k,2)} -$$
$$\tilde{T}^{-(k+1)}M^{r(k+1,1)-1}\left(M^{r(k,1)} - M^{r(k+1,1)}\right)Q^{(k,1)}.$$

As it can be seen, this system does not contain the inverse matrices $\tilde{T}^{\pm(k)-1}$.

If elements of matrices $M^{r(k,i)}$ vary sufficiently slowly with $k$, then the differences $\left|M^{r(k+1,i)}_{s,m} - M^{r(k,i)}_{s,m}\right|$ are the small values and we can get approximate equations by neglecting them.

The matrices $M^{r(k,i)}$ are the solutions of the same quadratic matrix equations

$$\tilde{T}^{-(k)} + \left(\tilde{T}^{+(k)}M^{r(k,i)} - I\right)M^{r(k,i)} = 0. \tag{16}$$

For homogeneous waveguide $\tilde{T}^{-(k)} = \tilde{T}^{+(k)}$, so if the waveguide structure varies sufficiently slowly with $k$, then the differences between entries of these matrices is small.

$\tilde{T}^{-(k)}$ and $\tilde{T}^{+(k)}$ are non-defective and we can write their eigenvalue decomposition as

$$\tilde{T}^{+(k)} = U^{+(k)}\Theta^{+(k)}U^{+(k)-1}, \tag{17}$$
$$\tilde{T}^{-(k)} = U^{-(k)}\Theta^{-(k)}U^{-(k)-1} = \left(U^{+(k)} + \Delta U^{-(k)}\right)\left(\Theta^{+(k)} + \delta\Theta^{-(k)}\right)\left(U^{+(k)} + \Delta U^{-(k)}\right)^{-1}. \tag{18}$$

Since the matrices $\Delta U^{-(k)}$ and $\delta\Theta^{-(k)}$ are small, we can neglect by quadratic terms

$$\tilde{T}^{-(k)} = U^{-(k)}\Theta^{-(k)}U^{-(k)-1} = \tilde{T}^{+(k)} + \Delta_1\tilde{T}^{-(k)} + \Delta_2\tilde{T}^{-(k)}, \tag{19}$$

where

$$\Delta_1\tilde{T}^{-(k)} = U^{+(k)}\delta\Theta^{-(k)}U^{+(k)-1}, \tag{20}$$
$$\Delta_2\tilde{T}^{-(k)} = \Delta U^{-(k)}\Theta^{+(k)}U^{+(k)-1} - U^{+(k)}\Theta^{+(k)}U^{+(k)-2}\Delta U^{-(k)}. \tag{21}$$

It can be shown (see below) that $\|\Delta_2 \tilde{T}^{-(k)}\| \ll \|\Delta_1 \tilde{T}^{-(k)}\|^3$ and it is natural to make the assumption that we can neglect $\Delta_2 \tilde{T}^{-(k)}$, then

$$\tilde{T}^{-(k)} \approx \tilde{T}^{+(k)} + \Delta_1 \tilde{T}^{-(k)} = U^{+(k)} \Theta^{-(k)} U^{+(k)-1}. \tag{22}$$

With this approximation we can find the solutions of the quadratic matrix equations (16). Taking into account the decompositions (17) and (22), we can find that

$$M^{r(k,i)} = U^{+(k)} \Lambda^{r(k,i)} U^{+(k)-1}, \tag{23}$$

where $i = 1, 2$, $\Lambda^{r(k,i)} = diag(\lambda_1^{r(k,i)}, \lambda_2^{r(k,i)}, ...)$.

Then from (16) we get characteristic equations for $\lambda_s^{r(k,i)}$

$$\theta_s^{+(k)} \lambda_s^{r(k,i)2} - \lambda_s^{r(k,i)} + \theta_s^{-(k)} = 0, \tag{24}$$

the solutions of which are

$$\lambda_s^{r(k,i)} = \frac{1}{2\theta_s^{+(k)}} \mp \sqrt{\left(\frac{1}{2\theta_s^{+(k)}}\right)^2 - \frac{\theta_s^{-(k)}}{\theta_s^{+(k)}}}, \tag{25}$$

We can also transform the equations (15) without calculating the inverse matrices $\tilde{T}^{\pm(k)-1}$

$$Q^{(k+1,1)} = M_{Q_1}^{r(k)} Q^{(k,1)} = \left\{ M^{r(k+1,1)} + \tilde{M}^{r(k+1,1)} \left( M^{r(k,1)} - M^{r(k+1,1)} \right) \right\} Q^{(k,1)}, \tag{26}$$

$$Q^{(k+1,2)} = M_{Q_2}^{r(k)} Q^{(k,2)} = \left\{ M^{r(k+1,2)} - \tilde{M}^{r(k+1,2)} \left( M^{r(k,2)} - M^{r(k+1,2)} \right) \right\} Q^{(k,2)}, \tag{27}$$

where $\tilde{M}^{r(k,i)} = U^{+(k)} \tilde{\Lambda}^{r(k,i)} U^{+(k)-1}$, $\tilde{\Lambda}^{r(k,i)} = diag\left(\tilde{\lambda}_1^{r(k,i)}, \tilde{\lambda}_2^{r(k,i)}, ...\right)$, $\tilde{\lambda}_s^{r(i)} = \lambda_s^{r(k,i)} \left( \lambda_s^{r(k,1)} - \lambda_s^{r(k,2)} \right)^{-1}$. We will call $Q^{(k,1)}$ the forward solution ($|\lambda_s^{r(k,1)}| \xrightarrow{s \to \infty} 0$) and $Q^{(k,2)}$ as the backward ($|\lambda_s^{r(k,2)}| \xrightarrow{s \to \infty} \infty$).

Equation (27) gives growing solutions when moving in the positive direction of the waveguide axis; therefore, it is necessary to calculate the field by moving in the negative direction. For this, we need to find a matrix inverse to $M_{Q_2}^{r(k)}$

$$Q^{(k,2)} = M_{Q_2}^{r(k)-1} Q^{(k+1,2)}. \tag{28}$$

As the matrix $M_{Q_2}^{r(k)}$ contains the difference $\left( M^{r(k,2)} - M^{r(k+1,2)} \right)$, its inverse cannot be found by using the properties of eigenvalue decomposition. Numerical methods also cannot be used, since with $|\lambda_s^{r(k,2)}| \xrightarrow{s \to \infty} \infty$ their accuracy will be low.

To find the WKB equation for the backward component $Q^{(k,2)}$ in the reverse order we introduce a new decomposition and rewrite the condition (12) in the form

$$Q^{(k-1)} = M^{l(k,1)} \tilde{Q}^{(k,1)} + M^{l(k,2)} \tilde{Q}^{(k,2)}. \tag{29}$$

After making the same transformations as above, we get

$$\tilde{Q}^{(k+1,2)} = M_{Q_2}^{l(k)} \tilde{Q}^{(k,2)} = \left\{ M^{l(k+1,2)} - \tilde{M}^{(k+1,2)} \left( M^{l(k,2)} - M^{l(k+1,2)} \right) \right\} \tilde{Q}^{(k,2)}, \tag{30}$$

where $M^{l(k,i)} = U^{+(k)} \Lambda^{l(k,i)} U^{+(k)-1}$, and $\Lambda^{l(k,i)} = diag(\lambda_1^{l(k,i)}, \lambda_2^{l(k,i)}, ...)$ and

$$\lambda_s^{l(k,i)} = \frac{1}{2\theta_s^{-(k)}} \mp \sqrt{\left(\frac{1}{2\theta_s^{-(k)}}\right)^2 - \frac{\theta_s^{+(k)}}{\theta_s^{-(k)}}}. \tag{31}$$

Matrices $M^{l(k,i)}$ and $M^{r(k,i)}$ are inverse to each other $M^{l(k,i)} M^{r(k,i)} = I$.

It was shown that in the case of difference equation given above procedure gives the multipliers $M_Q^{(k)}$ which coincide with the solutions of the same Riccati equation with error of $O(\varepsilon^2)$ [14]. So, we can expect that this is true for the matrix equations and put $\tilde{Q}^{(k,2)} = Q^{(k,2)}$. Finally, our WKB system of equations take the form

$$Q^{(k+1,1)} = M_{Q_1}^{r(k)} Q^{(k,1)} = \left\{ M^{r(k+1,1)} + \tilde{M}^{r(k+1,1)} \left( M^{r(k,1)} - M^{r(k+1,1)} \right) \right\} Q^{(k,1)}, \tag{32}$$

$$Q^{(k,2)} = M_{Q_2}^{l(k)} Q^{(k+1,2)} = \left\{ M^{r(k,2)-1} - \tilde{M}^{r(k,1)} \left( M^{r(k,2)-1} - M^{r(k+1,2)-1} \right) \right\} Q^{(k+1,2)}. \tag{33}$$

These equations have such solutions

---

[3] We will use the Frobenius norm $\|M\|_F = \sqrt{\sum_{i,j} M_{i,j}^2}$

$$Q^{(k+1,1)} = M_{Q_1}^{r(k)} \times ... \times M_{Q_1}^{r(N_C)} Q^{(N_C,1)} = \prod_{k=N_C}^{k} M_{Q_1}^{r(k)} Q^{(N_C,1)}$$

$$Q^{(k,2)} = M_{Q_2}^{l(k)} \times ... \times M_{Q_2}^{l(N_{REZ}-N_C)} Q^{(N_{REZ}-N_C+1,2)} = \prod_{k=N_{REZ}-N_C}^{k} M_{Q_2}^{l(k)} Q^{(N_{REZ}-N_C+1,2)}$$

(34)

Using these solutions, the system of linear equations (size $(N_Z \times N_{REZ}) \times (N_Z \times N_{REZ})$) that describe the chain of $N_{REZ}$ resonators ($Q \in C^{N_Z}$) [10]

$$T^{(Q_1)}Q^{(1)} + T^{(Q_2)}Q^{(2)} = Z^{Q1},$$
$$T^{(k)}Q^{(k)} = T^{+(k)}Q^{(k+1)} + T^{-(k)}Q^{(k-1)}, \quad k = 2,...,N_{REZ}-1,$$
$$T^{(Q_{NREZ}-1)}Q^{(N_{REZ}-1)} + T^{(Q_{NREZ})}Q^{(N_{REZ})} = 0,$$

(35)

can be transformed into a new system with much smaller size $(N_Z \times 2N_C) \times (N_Z \times 2N_C)$ as $N_C \ll N_{REZ}$.

$$T^{(Q_1)}Q^{(1)} + T^{(Q_2)}Q^{(2)} = Z^{Q1},$$
$$T^{(k)}Q^{(k)} = T^{+(k)}Q^{(k+1)} + T^{-(k)}Q^{(k-1)}, \quad k = 2,...N_C-2,$$
$$T^{-(N_c-1)}Q^{(N_c-2)} - T^{(N_c-1)}Q^{(N_c-1)} + T^{+(N_c-1)}Q^{(N_c,1)} + T^{+(N_c-1)} \prod_{k=N_{REZ}-N_c}^{N_c} M_{Q_2}^{l(k)} Q^{(N_{REZ}-N_c+1,2)} = 0,$$
$$T^{-(N_c)}Q^{(N_c-1)} - T^{(N_c)}Q^{(N_c,1)} - T^{(N_c)} \prod_{k=N_{REZ}-N_c}^{N_c} M_{Q_2}^{l(k)} Q^{(N_{REZ}-N_c+1,2)} + T^{+(N_c)} M_{Q_1}^{r(N_c)} Q^{(N_c,1)} +$$
$$T^{+(N_c)} \prod_{k=N_{REZ}-N_c}^{N_c+1} M_{Q_2}^{l(k)} Q^{(N_{REZ}-N_c+1,2)} = 0,$$
$$T^{-(N_{REZ}-N_c+1)} \prod_{k=N_C}^{N_{REZ}-N_c-1} M_{Q_1}^{r(k)} Q^{(N_c,1)} - T^{(N_{REZ}-N_c+1)} \prod_{k=N_C}^{N_{REZ}-N_c} M_{Q_1}^{r(k)} Q^{(N_c,1)} - T^{(N_{REZ}-N_c+1)} Q^{(N_{REZ}-N_c+1,2)} +$$
$$+T^{-(N_{REZ}-N_c+1)} M_{Q_2}^{l(N_{REZ}-N_c)} Q^{(N_{REZ}-N_c+1,2)} + T^{+(N_{REZ}-N_c+1)} Q^{(N_{REZ}-N_c+2)} = 0,$$
$$T^{-(N_{REZ}-N_c+2)} \prod_{k=N_C}^{N_{REZ}-N_c} M_{Q_1}^{r(k)} Q^{(N_c,1)} + T^{-(N_{REZ}-N_c+2)} Q^{(N_{REZ}-N_c+1,2)} - T^{(N_{REZ}-N_c+2)} Q^{(N_{REZ}-N_c+2)} +$$
$$T^{+(N_{REZ}-N_c+2)} Q^{(N_{REZ}-N_c+3)} = 0,$$
$$T^{(k)}Q^{(k)} = T^{+(k)}Q^{(k+1)} + T^{-(k)}Q^{(k-1)}, \quad k = N_{REZ}-N_c+3,...N_{REZ}-1,$$
$$T^{(Q_{NREZ}-1)}Q^{(N_{REZ}-1)} + T^{(Q_{NREZ})}Q^{(N_{REZ})} = 0$$

(36)

Decomposition (11)-(12) and the solutions (34) were used for $k = N_C,....,N_{REZ}-N_C+1$ as the approximate method is not applicable for the ending cells. We assumed that $N_C = 10$.

We have used this system to study the properties of the proposed approximate approach.

## 2. SIMULATION RESULTS

Let us consider the chain of $N_{REZ}$ resonators[4] coupling through cylindrical openings in the walls of finite thickness. On each side, the chain is connected to a semi-infinite cylindrical waveguide. We will consider only axially symmetric fields with $E_z, E_r, H_\varphi$ components (TM). Time dependence is $\exp(-i\omega t)$. We also suppose that in the left semi-infinite waveguide the $TM_{0,1}$ eigen wave with a unitary amplitude propagates towards the considered chain. We will consider the chain without dissipation of electromagnetic energy in walls.

We will investigate the characteristics of a smooth transition between two disk waveguides with parameters: $a_I = 1.3810$ cm, $b_I = 4.1618$ cm, $\beta_{g,I} = v_{g,I}/c = 0.02$, $a_{II} = 1.0200$ cm, $b_{II} = 4.0785$ cm, $\beta_{g,II} = 0.0062$, $d_I = d_{II} = 2.9147$ cm, $t_I = t_{II} = 0.5842$ cm, $D = d_I + d_{II} = \lambda_0/3$, $\omega_0 = 2\pi 2.856$ GHz and the phase shift per cell $\varphi_I = \varphi_{II} = 2\pi/3$[5]. The aperture and resonator radii change according the formulas (see [10])

$$a_k = 0.6745 + \sqrt{-0.0506 + 27.4745 \beta_{g,k}}$$ (37)

$$b_k = 0.1576 a_k^2 - 0.1476 a_k + 4.0652.$$ (38)

where $k = 3,..., N_{REZ}-2$ and

---
[4] Segment of the inhomogeneous disk-loaded waveguide
[5] These parameters correspond the initial and final cells of the SLAC 3m section

$$\beta_{g,k} = \frac{\beta_{g,I} + \beta_{g,II}}{2} + \frac{\beta_{g,I} - \beta_{g,II}}{2} \frac{\arctan\{\alpha(k-k_0)\}}{\arctan\{\alpha(3-k_0)\}},. \tag{39}$$

The formulas (37) and (38) define the sizes of homogeneous disk-loaded waveguide which has a phase shift $\varphi = 2\pi/3$ and a group velocity $\beta_{g,k}$ [10].

The dependences of the geometric dimensions of the two transitions with $\alpha = 0.03$ and $\alpha = 0.1$ (denoted as 1 and 2) are shown in Figure 3 and Figure 4.

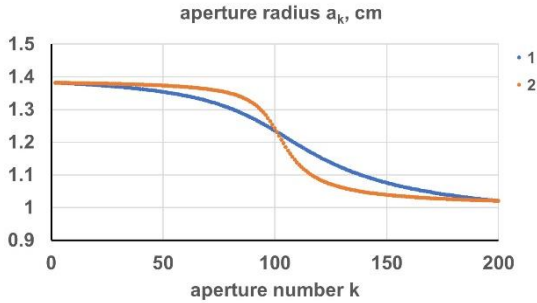
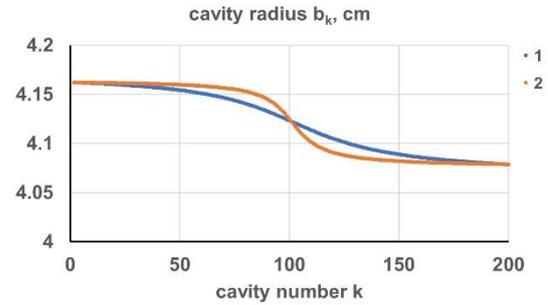

Figure 3 Aperture radius $a_k$ as function of cavity number (for the $k$-th resonator the left aperture has a number $k$, the right aperture -$(k+1)$) for two transitions: 1- $\alpha = 0.03$, 2 - $\alpha = 0.1$

Figure 4 Resonator radius $b_k$ as function of cavity number for two transitions: 1- $\alpha = 0.03$, 2 - $\alpha = 0.1$

The distributions of the longitudinal component of the electric field for the two transitions, calculated with using the CASCIE code, are presented in Figure 5 and Figure 6.

From Figure 6Figure 1 it follows that a constant phase shift appears in the case of choosing the local dimensions of an inhomogeneous waveguide equal corresponding sizes of homogeneous structures [10].

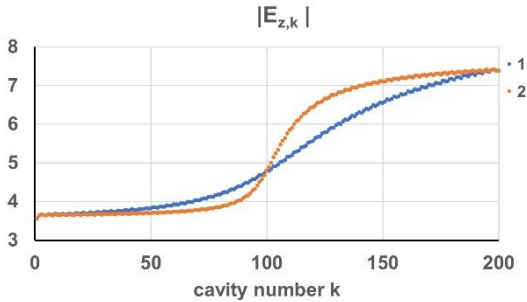
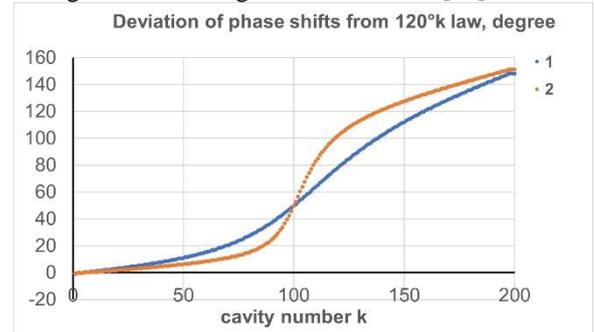

Figure 5 The amplitude of the longitudinal electric field at the centers of resonators for two transitions: 1- $\alpha = 0.03$, 2 - $\alpha = 0.1$

Figure 6 The phase of the longitudinal electric field at the centers of resonators for two transitions: 1- $\alpha = 0.03$, 2 - $\alpha = 0.1$

In section 2 we made two important assumptions. The first assumption is $\left\|\Delta_2 \tilde{T}^{-(k)}\right\| \ll \left\|\Delta_1 \tilde{T}^{-(k)}\right\|$. From Figure 7, in which the norm of matrices $\Delta_1 \tilde{T}^{-(k)}$ and $\Delta_2 \tilde{T}^{-(k)}$ are represented for different cells, follow that this assumption is true even for the case of a rapid change in geometric parameters ($\alpha = 0.5$).

The second assumption is $\tilde{Q}^{(k,2)} = Q^{(k,2)}$. We can test this assumption by considering a chain with total reflection at the right end. If the assumption is correct, we should get $\left|Q^{(k,1)}\right| \approx \left|Q^{(k,2)}\right|$ and, as a result, $\left|E_{z,k}^+(d_k/2)\right| \approx \left|E_{z,k}^-(d_k/2)\right|$, where $E_{z,k}^\pm(r=0, d_k/2) = \sum_s Q_s^{\left(\substack{1\\2},k\right)}$ - the longitudinal components of electric field. From Figure 8 it follows that this assumption is valid only for the case of a slow change in the geometrical parameters of the waveguide. Detailed analysis shows that in the standing wave regime this assumption gives the greatest error in numerical calculations.

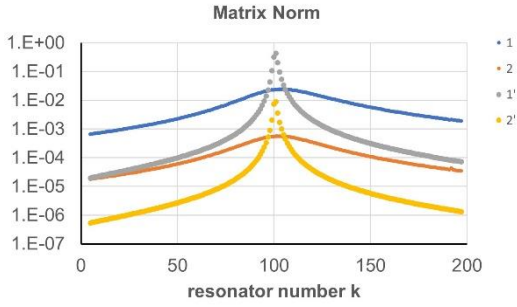 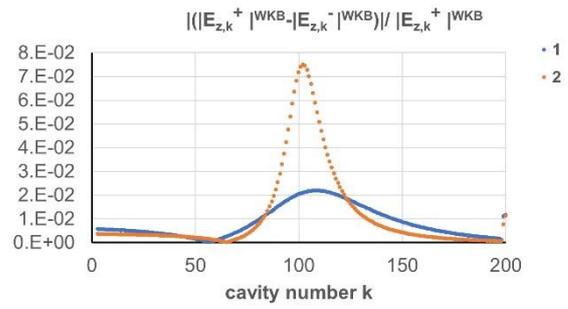

Figure 7 Norms of the matrices $\Delta_1 \tilde{T}^{-(k)}$ (1 and 1') and $\Delta_2 \tilde{T}^{-(k)}$ (2 and 2') for two transitions: 1,2- $\alpha = 0.05$, 1',2' - $\alpha = 0.5$

Figure 8 Difference between the forward solution and the backward solution with shorting at the end for two transitions: 1- $\alpha = 0.03$, 2 - $\alpha = 0.1$

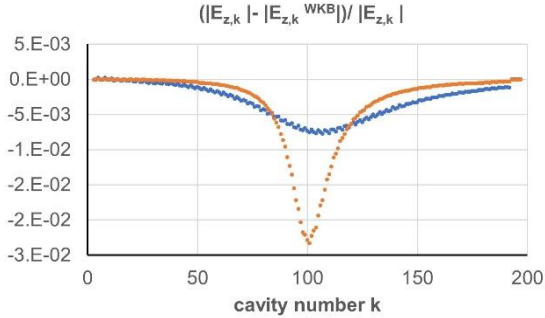 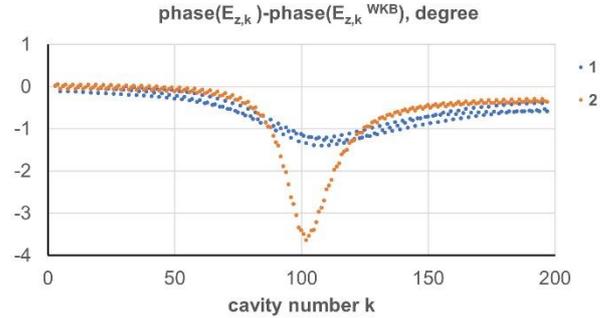

Figure 9 Difference between the solution of equations (2)( CASCIE) and the solution of the WKB equations (32) for two transitions: 1- $\alpha = 0.03$, 2 - $\alpha = 0.1$

Figure 10 Difference between the solution of equations (2)( CASCIE) and the solution of the WKB equations (32) for two transitions: 1- $\alpha = 0.03$, 2 - $\alpha = 0.1$

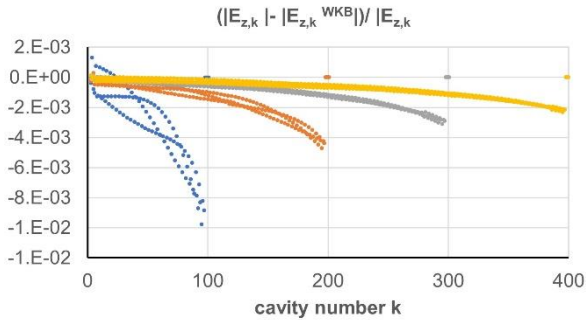 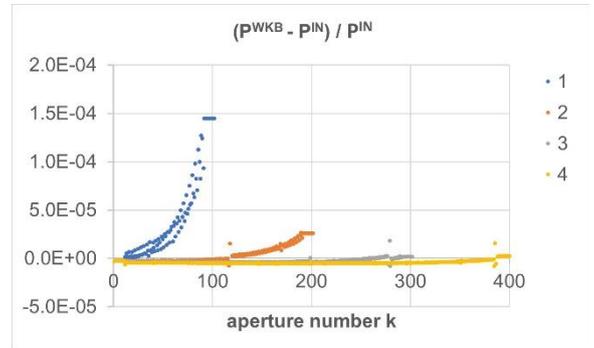

Figure 11 Difference between the solution of equations (2) ( CASCIE) and the solution of the WKB equations (32) for transitions with $\alpha = 1E-6$ and $N_{REZ} = 101$ (1), 201 (2), 301 (3), 401 (4)

Figure 12 The difference in power flows through the k-th and first apertures for transitions with $\alpha = 1E-6$ and $N_{REZ} = 101$ (1), 201 (2), 301 (3), 401 (4)

The accuracy of the considered approximate method can be estimated by comparing the solution of the approximate system of equations (36) with the solution of the general system (35). Results of such comparing are presented in Figure 9 and Figure 10. For the homogeneous waveguide there is no difference between the solutions of system (36) and system (35) (an error is less than $10^{-8}$) .

For accelerator technology, inhomogeneities with a linear decrease in the parameter $\beta_{g,k}$ (39) are of great interest. It follows from (39) that $\alpha \to 0$ corresponds to this case. We remind that the initial and final cells of the considered transitions correspond to the initial and final cells of the SLAC 3m section. From Figure 11 and Figure 12 it follows that the accuracy of the proposed approximate approach for sections with "group velocity" gradients, similar to the SLAC section, is about 1%. Using the proposed WKB method for describing sections with a larger gradient [15,16] will give larger errors - (3-4)%.

## CONCLUSIONS

The results of the development of an approximate approach, which can be considered as an analogue of the WKB method, are presented. This approach gives possibility to divide the electromagnetic field into forward and backward components and simplify the analysis of the field characteristics, especially the phase distribution. The accuracy of this method was estimated by comparing the solution of the approximate system of equations with the solution of the general system of equations. For this, a special code was written that combines the proposed approach with the more accurate one developed earlier. The results of this comparison showed that the accuracy of the proposed approximate approach for sections with "group velocity" gradients, similar to the SLAC section, is about 1%. The proposed approach and the results obtained are especially important in the case of studying the energy spread and its minimum, which can be achieved using a low-level RF feedback system.

## ACKNOWLEDGEMENTS

The author would like to thank David Reis and Valery Dolgashev for their support and interest in work.